\documentclass[pra,preprint,showpacs,endfloats]{revtex4}

\usepackage{epsfig}
\usepackage{graphicx}
\usepackage{bm}
\usepackage{amsmath}
\usepackage{dcolumn}

\newcommand{\rcm}{\mbox{cm$^{-1}$}}
\newcommand{\Sch}{Schr\"{o}dinger}
\newcommand{\Sstate}{$^1\Sigma^+$}
\newcommand{\Pstate}{$^1\Pi$}

\begin{document}

\title{On the B\Pstate\ state of NaCs}

\author{A. Grochola, P. Kowalczyk}
\thanks{Corresponding author}
\email {pfkowal@fuw.edu.pl} \affiliation{Institute of Experimental
Physics, Warsaw University, ul. Ho\.za 69, 00-681~Warsaw, Poland}
\author{W. Jastrzebski}
\affiliation{Institute of Physics, Polish Academy of Sciences,
Al.Lotnik\'{o}w 32/46, 02-668~Warsaw, Poland}

\date{\today}

\begin{abstract}
The B\Pstate\ $\leftarrow$ X\Sstate\ system of NaCs molecule is
investigated experimentally by polarization labelling
spectroscopy technique. The inverted perturbation approach method
is used to construct the potential energy curve of the B~\Pstate\
state, providing accurate eigenenergies for vibrational levels $v
= 0 - 15$.
\end{abstract}


 \maketitle


The NaCs molecule belongs to the least known heteronuclear alkali
dimers. Only the two lowest electronic states, X\Sstate\ and
a$^3\Sigma^+$, have been studied thoroughly and with high
accuracy \cite{1,2}. In addition, three excited states available
by optical excitation from the ground state, B(1)\Pstate,
D(2)\Pstate\ and 3\Pstate, have been observed and partially
characterized \cite{1,3,4,5}. Our present study is focused on the
B\Pstate\ state which is of considerable interest because of its
involvement in photoassociation processes \cite{6,7}. When
observing excitation spectra of the B\Pstate\ $\leftarrow$
X\Sstate\ band system of NaCs we found that the previously
published potential energy curve of the B\Pstate\ state \cite{5}
provides eigenenergies of rovibrational levels which are off by
up to 5~\rcm\ comparing the experimental ones. In this
communication we present an improved characterization of the
lower part of the B\Pstate\ state. A work on full description of
this state is currently in progress in our laboratory.

The experimental arrangement and the employed V-type
optical-optical double resonance polarization labelling
spectroscopy technique were very similar to that described in our
previous papers \cite{8,9}. NaCs molecules were produced in a
heat-pipe oven, heated to about 940~K and filled with 4~mbar of
helium buffer gas. Our method requires two independent pump and
probe lasers. The copropagating laser beams were crossed in the
centre of the heat-pipe. The linearly or circularly polarized
pump beam induced transitions from the ground state of NaCs to
the investigated B\Pstate\ state, while the probe beam was set at
fixed wavelengths resonant with known transitions in the
D\Pstate\ $\leftarrow$ X\Sstate\ system \cite{1}, thus labelling
known rovibrational levels in the X\Sstate. As a pump laser
served the optical parametric oscillator with an amplifier
(Sunlite Ex, Continuum) pumped with an injection seeded Nd:YAG
laser (Powerlite 8000). This system provided laser pulses with
energy of a few mJ and spectral width approximately 0.1~\rcm,
tuneable in the present experiment in the $625 - 653$~nm range.
The wavelength of the pump laser was controlled and calibrated to
better than 0.1~\rcm\ by measuring two additional signals:
optogalvanic spectrum of argon and transmission fringes of a
Fabry-P\'{e}rot interferometer 0.5~cm long. As the probe laser we
used a home built dye laser pumped synchronously with the same
Nd:YAG laser and operated on Coumarin 153 dye. The laser
frequency was measured with HighFinesse WS-7 wavemeter. Crossed
polarizers were placed at both sides of the oven in the path of
the probe beam. At these frequencies, at which transitions
induced by the pump beam shared the same lower level with the
probe transition, some of the probe light passed through the
analyzer. This residual beam was monitored with a photomultiplier
tube connected to the boxcar averager (Stanford Research Systems,
SR250). Thus obtained excitation spectra of the B\Pstate\
$\leftarrow$ X\Sstate\ system of NaCs were stored in a computer
together with the reference spectra of argon and frequency
markers of the interferometer.

\begin{figure}
  \centering
\epsfig{file=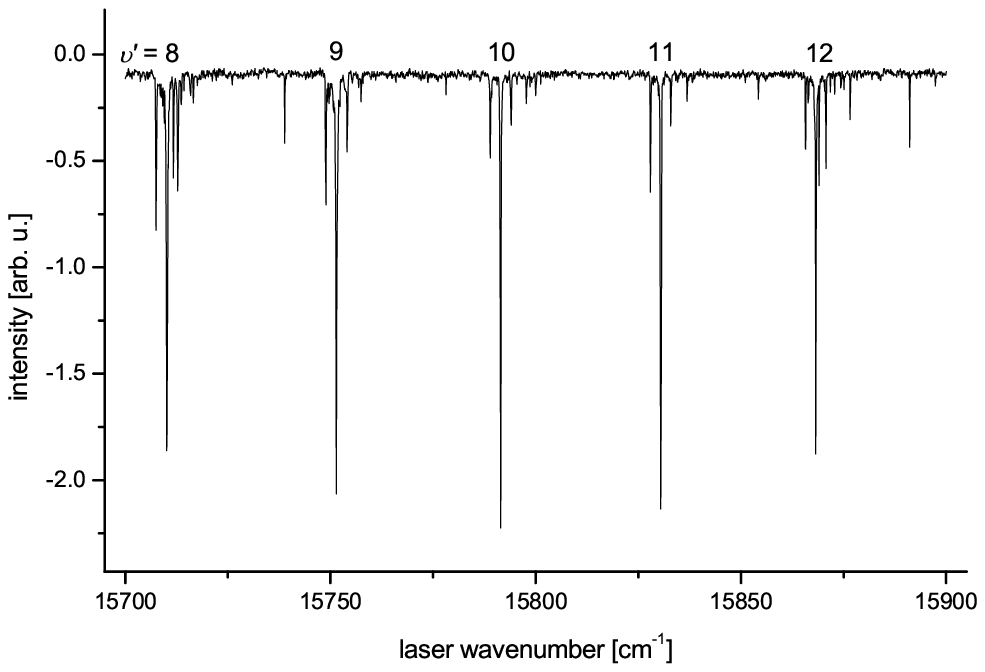,width=\linewidth}
 \caption {A portion of the polarization spectrum of NaCs. The assigned
 progression corresponds to transitions B\Pstate\ ($v'$, $J' = J''$, $J''\pm1$)
 $\longleftarrow$ X\Sstate\ ($v'' = 0$, $J'' = 36$) labelled by the probe laser
 set at the wavenumber 18348.45~\rcm.}
  \label{fig1}
   \end{figure}

An example of the recorded spectrum of NaCs is shown in Figure
\ref{fig1}. The highly accurate molecular constants of the ground
X\Sstate\ state \cite{1} and energy levels computed from the
existing B\Pstate\ state potential curve \cite{5} provided
unambiguous assignment of the observed lines. However, we found
that whereas the experimental positions of the lowest ($v = 0 -
5$) vibrational levels in the B state followed precisely the
predicted ones, the deviation grew systematically for higher
vibrational numbers, in some cases exceeding 5~\rcm. This
observation justified new determination of the excited state
potential. In view of a complex pattern of $\Lambda$ doubling in
the B\Pstate\ state pointed out in the previous work \cite{5} and
confirmed by observation of irregular spacing between the
relevant $P$, $Q$ and $R$ lines in the spectra, we confined the
further analysis to $Q$ lines only, corresponding to transitions
to $f$-parity levels in the upper state.

\begin{figure}
  \centering
\epsfig{file=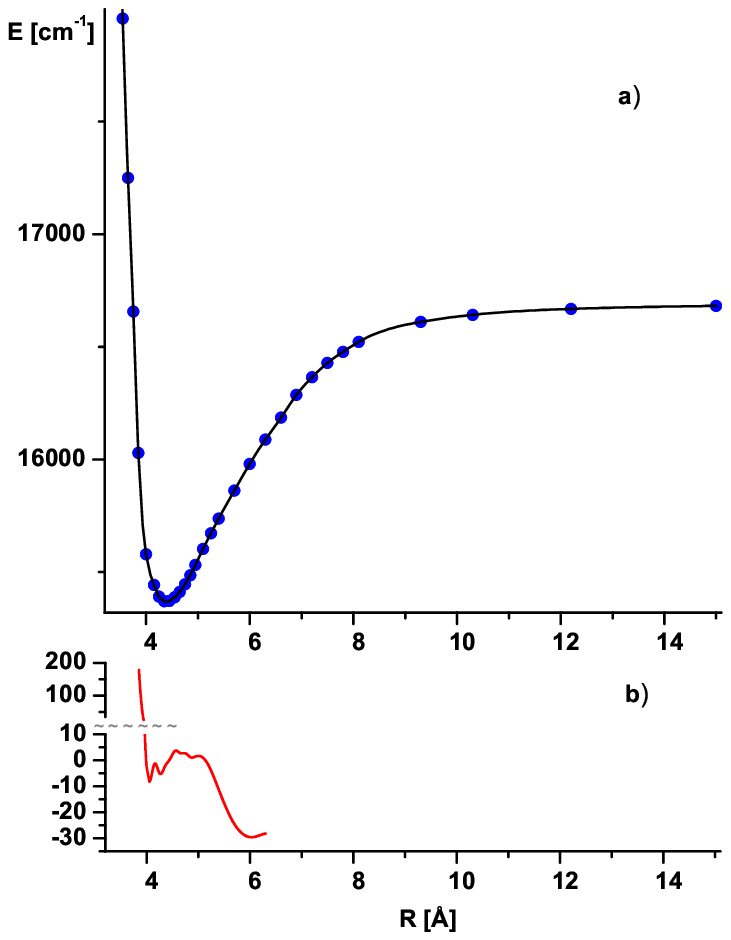,width=\linewidth}
 \caption {(a) The IPA potential energy curve of the B\Pstate\ state in NaCs;
 (b) Difference in energy between the presently determined potential
 and that given by Zaharova et al. \cite{5}, shown in the region of reliability
 of our curve.}
  \label{fig2}
   \end{figure}

The recorded spectra provided information about $185$
rovibrational $f$-parity levels in the B\Pstate\ state. We
identified levels with vibrational quantum numbers $v$ between 0
and 15 and rotation quantum numbers $J$ in the range $19 - 51$.
The measured wavenumbers of spectral lines were converted to
energies of the B\Pstate\ state levels referred to the bottom of
the X\Sstate\ state potential well using the ground state Dunham
coefficients of Diemer \textit{et al}. \cite{1}. The potential
curve of the B state was constructed with the pointwise inverted
perturbation approach (IPA) method \cite{10} by ensuring that the
rovibronic energies calculated from the potential curve matched
the experimental energies. The final curve is defined by 33
points (Table \ref{pot} and Figure \ref{fig2}) which are
interpolated using a natural spline algorithm \cite{11} to
generate a numerical mesh of 9000 points between 3 and 15~\AA. It
reproduces the measured term energies of vibrational levels in
the B\Pstate\ state with an r.m.s. error of 0.17~\rcm. This curve
differs significantly from the one published previously by
Zaharova et al. \cite{5}, as displayed in Figure \ref{fig2}. The
error analysis shows that the observed levels determine the
potential curve reliably only between 3.85 and 6.3~\AA. Points
outside this range are supplied here only to ensure proper
boundary conditions for solving the \Sch\ equation.

\begin{table}
\caption{ Parameters defining the rotationless IPA potential
energy curve of the B\Pstate\ state in NaCs. } \label{pot}
\vspace*{0.5cm}
\begin{tabular*}{1.0\columnwidth}{@{\extracolsep{\fill}}cccc}
 \hline
$R$ [\AA] & $U$ [\rcm] & $R$ [\AA] & $U$ [\rcm]  \\
 \hline
       3.00  & 29008.4143 & 5.10  & 15602.4424  \\
       3.30  & 21485.8095 & 5.25  & 15671.5661  \\
       3.45  & 19029.1720 & 5.40  & 15737.4482  \\
       3.55  & 17957.5184 & 5.70  & 15862.3295  \\
       3.65  & 17250.4519 & 6.00  & 15980.2352  \\
       3.75  & 16656.9477 & 6.30  & 16088.2737  \\
       3.85  & 16028.9352 & 6.60  & 16185.4479  \\
       4.00  & 15579.1650 & 6.90  & 16286.3384  \\
       4.15  & 15443.4526 & 7.20  & 16365.2600  \\
       4.25  & 15390.9772 & 7.50  & 16428.0184  \\
       4.35  & 15370.8956 & 7.80  & 16478.1001  \\
       4.45  & 15371.1314 & 8.10  & 16521.9956  \\
       4.55  & 15387.6360 & 9.30  & 16609.9880  \\
       4.65  & 15412.1144 & 10.30 &  16641.3947  \\
       4.75  & 15446.3142 & 12.20 &  16669.0511  \\
       4.85  & 15485.5787 & 15.00 &  16682.1325  \\
       4.95  & 15530.9193 \\
\hline
\end{tabular*}
\end{table}

\begin{acknowledgments}
This work has been funded in part by Grant No. N202 103 31/0753
from the Polish Ministry of Science and Higher Education.
\end{acknowledgments}

\end{document}